\documentclass[]{article}
\usepackage[applemac]{inputenc}
\usepackage[T1]{fontenc}
\usepackage[dvips]{graphicx}
\usepackage{psfrag}
\usepackage{pstricks}
\usepackage{acronym}
\usepackage{latexsym}

\usepackage[english]{babel}
\usepackage{epsfig,array,amsmath,amssymb,float}
\usepackage{xspace,calc,multirow,rotating,theorem}
\usepackage[numbers, super]{natbib} 
 \usepackage{setspace}

\usepackage{longtable,calc,multirow}
\DeclareMathAlphabet{\mathbfeul}{U}{eur}{b}{n}
\DeclareMathAlphabet{\matheul}{U}{eur}{m}{n}
\DeclareMathAlphabet{\mathbfeus}{U}{eus}{b}{n}
\DeclareMathAlphabet{\matheus}{U}{eur}{m}{n}
\DeclareMathAlphabet{\mathbfzc}{OML}{pzc}{b}{n}
\DeclareMathAlphabet{\mathzc}{OT1}{pzc}{m}{it}
\DeclareMathAlphabet{\mathag}{OT1}{pag}{m}{n}
\DeclareMathAlphabet{\mathbfeuf}{U}{euf}{b}{n}
\DeclareMathAlphabet{\matheuf}{U}{euf}{m}{n}

\newcommand{\qedsymbol}{\rule{1ex}{1ex}}
\newcommand{\qed}{\mbox{}\hfill\qedsymbol}

\newcommand{\bMatsmall}[1]{\left[\begin{smallmatrix}#1\end{smallmatrix}\right]}

\newcommand {\tr} {\rm\scriptscriptstyle T}
\newcommand {\grad}[2]{\frac{\partial#1}{\partial#2}}

\newcommand{\beali}[1]{\begin{equation}\begin{aligned}#1\end{aligned}\end{equation}}

\newcommand{\bea}[1]{\begin{eqnarray} #1 \end{eqnarray}}

\newcommand{\bMat}[1]{\begin{bmatrix}#1\end{bmatrix}}

\newcommand{\inn}{{\scriptscriptstyle in}}
\newcommand{\reac}{{\scriptscriptstyle r}}

\newcommand{\inv}{{\scriptscriptstyle iv}}

\newcommand{\out}{{\scriptscriptstyle out}}

\newcommand {\eqrefn}{Eq.~\eqref}
\newcommand{\rank}[1]{\operatorname{rank}\left(#1\right)}  
  \newcommand{\cbf}{\mathbf{c}}
  \newcommand{\fbf}{\mathbf{f}}

 \newcommand{\nbf}{\mathbf{n}} 
  \newcommand{\rbf}{\mathbf{r}}
  \newcommand{\ubf}{\mathbf{u}}
 \newcommand{\wbf}{\mathbf{w}} \newcommand{\xbf}{\mathbf{x}}
 \newcommand{\zbf}{\mathbf{z}}

\newcommand{\betabf}{\boldsymbol{\beta}}

  \newcommand{\Ibf}{\mathbf{I}}
\newcommand{\Jbf}{\mathbf{J}}  
\newcommand{\Mbf}{\mathbf{M}} \newcommand{\Nbf}{\mathbf{N}} 
 \newcommand{\Qbf}{\mathbf{Q}} 
  
 \newcommand{\Wbf}{\mathbf{W}} 
  
\newcommand{\ones}{\boldsymbol{1}}
\newcommand{\zeros}{\boldsymbol{0}}\newcommand{\Rbb}{\mathbb{R}}
{\theoremstyle{break}\theoremheaderfont{\normalfont\bfseries}}
{\theoremstyle{plain}\theorembodyfont{\normalfont\rmfamily}}
{\theoremstyle{break}\newtheorem{defi}{Definition}\theoremheaderfont{\normalfont\bfseries}
{\theoremstyle{break}\theoremheaderfont{\normalfont\bfseries}
{\theoremstyle{break}\theoremheaderfont{\normalfont\bfseries}
{\theoremstyle{break}\theoremheaderfont{\normalfont\bfseries}
{\theoremstyle{break}\theoremheaderfont{\normalfont\bfseries}
{\theoremstyle{break}\theoremheaderfont{\normalfont\bfseries}}
{\theoremstyle{break}\theoremheaderfont{\normalfont\bfseries}
{\theoremstyle{break}\newtheorem{proposition}{Proposition}\theoremheaderfont{\normalfont\bfseries}

{\theoremstyle{break}\newtheorem{corollary}{Corollary}\theoremheaderfont{\normalfont\bfseries}

{\theoremstyle{break}\theorembodyfont{\normalfont\rmfamily}}
{\theoremstyle{break}\theorembodyfont{\normalfont\rmfamily}}

\newcommand {\vlightrule}{\kern1ex\vrule width0.2pt\kern1ex}
\definecolor{GreenByNirav}{rgb}{0.2,0.5,0.21}

\newtheorem{lemma}{Lemma}

\def\endkeywords{\vspace{0.6em}\par\if@twocolumn\else\endquotation\fi
    \normalsize\rm}

\begin{document}
\title{On Cooperative Behavior of Open Homogeneous Chemical Reaction Systems in the Extent Domain}
\author{
        \begin{tabular}[t]{cc}
     Nirav Bhatt$^{1}$,\footnote{Corresponding author. E-mail: {\tt niravbhatt@iitm.ac.in}}     & Sriniketh Srinivasan$^{2}$
        \end{tabular}	\\
	$^{1}$Systems \& Control Group,\\
	Indian Institute of Technology, Madras,\\
	Chennai - 600036, India \\
	$^{2}$Laboratoire d'Automatique,\\ \'{E}cole Polytechnique Fédérale de Lausanne,  \\          CH-1015 Lausanne, Switzerland\\
        }

\maketitle

\date{}

\begin{abstract}
Material balance equations describe the dynamics of the species in open reaction systems  and contain information regarding reaction topology, kinetics and operation mode. 
For reaction systems, the state variables (the numbers of moles, or concentrations) have recently been transformed into decoupled reaction
variants (extents of reaction), and reaction invariants (extents of flow) (Amrhein et al., AIChE Journal,
2010).  This paper analyses the conditions under which an open homogeneous reaction system is  cooperative in the extents domain. Further, it is shown that the dynamics of the  extents of flow  exhibit cooperative behavior.  Further, we provide the conditions under which the dynamics of the extents of reaction exhibit cooperative behavior. Our results provide physical insights into cooperative and competitive nature of the underlying   reaction system in the presence of material exchange with surrounding (i.e., inlet and outlet flows). The results of the article are demonstrated via  examples.  
\end{abstract}
\begin{keywords}
 Extent of reaction, Extents of flow, Open reaction systems, Cooperative Systems, Competitive Systems
\end{keywords}

\section{Introduction}
Chemical reaction systems are present in various spheres of our lives. They occur in  fields such as in basic sciences, chemistry and biology, and in applied sciences, chemical engineering and biotechnology. A quantitative and qualitative analysis  provides a better understanding of the underlying reaction system. In process industries, such  analysis is useful for smooth operation of these systems  via model-based monitoring, control and optimization. On the other hand, in  systems biology, it helps to understand functionalities of different biological networks under various conditions \cite{AngeliLS10}. In this paper, we investigate cooperative property of  open homogeneous reaction systems.

 For a \textit{single-phase homogeneous reaction system},  the states (concentrations or the numbers of moles) that evolve with reaction progress, labeled reaction variants, and the states that do not, labeled reaction invariants.   However, the reaction variants and invariants in the literature are merely mathematical quantities that do not have a direct physical meaning due to the couplings that exist between them \cite{Amrhein98}.  Recently,  a linear transformation to obtain decoupled and physical meaningful reaction and flow variants, called the extents of reaction and flow, respectively, has been proposed \cite{AmrheinBSB10}. The decoupling of variants in the extents domain has been exploited in various implications such as identification of reaction systems, state reconstruction/estimation, and minimal state representations \cite{SrinivasanBB13,BhattASMB12}.  
 
 The study of cooperative/competitive properties is an active area of research in monotonous  dynamical system theory. A system is referred to be cooperative/competitive systems, if the solutions of  system preserve monotonous order (increasing or decreasing). A rigorous definition of monotonous systems  can be found  in Smith, H.\cite{Smith95}. Further, monotonous  system theory can provide insight into dynamic behaviors of the underlying system based on the topology \cite{Smith95}. The dynamics  of chemical reaction systems can be modeled as the combination of topology (stoichiometric) and the interacting functions (reaction rates). Hence, the monotonous system theory has been applied to understand qualitative behaviour of these complex systems under various conditions in systems biology \cite{Sontag07}. On the other hand, these concepts are useful in developing interval estimators  in process control\cite{GouzeRH00}.  
 
 In this paper, we investigate  cooperative nature of  the system in the extents domain for open reaction systems. The main objective of the work is to investigate conditions for the underlying open reaction system being a cooperative one which can be easily verified from physical meaningful quantities in contrast to other authors in the literature \cite{AngeliLS10}. It will be shown  that the extents of flow always exhibit cooperative behavior. Further, we will identify the condition under which the extents of reaction (the embedded reactions) exhibit cooperative or competitive behavior. Moreover, it is shown that the extents of reaction can behave in  cooperative manner under certain experimental conditions. Further, the effect of operation mode on the embedded reactions will be investigated.  
 
 The paper is organized as follows. Some preliminaries on the linear transformation and basic definitions are given in Section 2. Section 3 provides main results of this paper. In this section, conditions under which the underlying reaction system being a cooperative system  in the extents domain are given. Section 4 demonstrates various concepts via two examples. Section 5 concludes the paper.   

%

\section{Preliminaries}

\subsection{Model of homogeneous reaction systems}

 In this section we write  material balance equations for an open non-isothermal homogeneous reaction system. 
\subsubsection*{\bf Mole Balance Equations} 

Consider a homogeneous reaction system with $S$ species living in a set of species $\mathcal{S}$ with $|\mathcal{S}|$=S. The reaction system consists of $R$  reactions, $p$ inlet streams, and one outlet stream. Then, the mole balance equations for the reaction system can be written as follows:
\beali{
\dot{\nbf}(t) =  \Nbf^{\tr} \rbf_v(t) + \Wbf_{\inn}\,\ubf_{\inn}(t)\, - \omega(t) \nbf(t),\,\, \nbf(0) = \nbf_0, \label{sys1}
 }
where $\nbf(t)$ is the $S$-dimensional vector of numbers of moles, $\rbf_v(t):=V(t)\, \rbf\big(\cbf(t)\big)\,$ with $V(t)$ the volume and $\rbf\big(\cbf(t)\big)$ the $R$-dimensional reaction rate vector, $\ubf_{\inn}(t) \geq \zeros_p$  the  $p$-dimensional inlet mass flowrate vector, $\omega(t):=u_{\out}(t)/m(t)\geq0$ the inverse of the reactor residence time with  $m(t)$ the mass of the reaction mixture and $u_{\out}(t)$ the outlet mass flowrate, $\Wbf_{\inn}=\Mbf_w^{-1}\check{\Wbf}_{\inn}$ the $S\times p$ inlet-composition matrix,  $\Mbf_w$ the $S$-dimensional diagonal matrix of molecular weights, $\check{\Wbf}_{\inn}=[\check{\wbf}^{1}_{\inn} \cdots \check{\wbf}^{p}_{\inn}]$ with  $\check{\wbf}^{j}_{\inn}$ being the $S$-dimensional vector of weight fractions of the $j$th inlet flow, 
 and $\nbf_0$ the $S$-dimensional vector of initial numbers of moles. $\Nbf$ the $R\times S$ stoichiometric matrix is constructed with the  stoichiometric coefficients $\nu_{ij}$, where $i = 1 \dots R$ and $j = 1 \dots S$ with $\nu_{ij} > 0$ for products and $\nu_{ij} < 0$ for reactants. \eqrefn{sys1} is an example of positive dynamic systems. 
\ \\
The species involved in a reaction can further be classified into reactant or product set based on the following definitions: 
\begin{defi}[Reactant Set]
For a given reaction $i$, the reaction set $\mathcal{S}_{i}^R$ is a subset of $\mathcal{S}$ containing species taking part in  the reaction $i$ with strictly negative stoichiometric coefficients.
$$ (i.e.) \, \,  \mathcal{S}_{i}^R \subset \mathcal{S} \qquad s.t \, \, \, \, \nu_{ij} < 0.  $$
$$ |\mathcal{S}_{i}^R| = S_{r,i} $$
\end{defi}
\begin{defi}[Product Set]
For a given reaction $i$, the product set $\mathcal{S}_{i}^P$ is a subset of $\mathcal{S}$ containing species taking part in the reaction $i$ with strictly positive stoichiometric coefficients. 
$$ (i.e.) \, \, \mathcal{S}_{i}^P \subset \mathcal{S} \qquad s.t \, \, \, \, \nu_{ij} > 0.  $$
$$ |\mathcal{S}_{i}^P| = S_{p,i} $$
\end{defi}

 \ \\
 \textbf{Assumption A1:} Here, it is assumed that a species cannot be both a reactant and a product in the same reaction, i.e. the reaction system is not autocatalytic. When Assumption A1 holds, we have, $\mathcal{S}_{i}^R \cap \mathcal{S}_{i}^{P} = \phi$ with $\phi$ denotes an empty set. \\

The flowrates $\ubf_{\inn}(t)$  and $u_{\out}(t)$ are considered as independent (input) variables in  \eqrefn{sys1}. 
The continuity equation (or total mass balance) is given by:
\beali{ \dot{m}(t) = \ones_{p}^{\tr}\ubf_{\inn}(t) - u_\out(t), \qquad\qquad m(0)=m_0, \label{eq:continuity}}
where $\textbf{1}_p$ is the $p$-dimensional vector filled with ones  and $m_0$ the initial mass. Note that the mass $m(t)$ can also be computed from the numbers of moles as
\beali{
m(t)=\ones_S^{\tr}\,\Mbf_w\,\nbf(t).\label{eq:m}
 }
The  volume $V(t)$ can be inferred from the mass, $V(t)=m(t)/\rho\big(\cbf(t),T(t)\big)$, with the density $\rho$ being a function of concentrations and temperature. The following three scenarios,  chemical reactions can exhibit (1) linearly-dependent stoichiometries, (2) linearly dependent reaction rates on a given time interval, and/or (3) time-varying stoichiometries. 
Without loss of generality, it is assumed that the $R$ reactions and the $p$ inlets are independent according to the definitions given next:
\begin{defi}[Independent reactions]
The $R$ reactions are said to be independent if (i) the rows of $\,\Nbf$ (stoichiometries) are linearly independent, i.e.\ $\rank{\Nbf}=R$, and (ii) there exists some finite time interval $t \in [t_0,\,t_1]$
for which the reaction rate profiles $\rbf(t)$ are linearly independent, i.e.\ $\betabf^{\tr}\rbf(t) = 0$ $\Leftrightarrow$ $\betabf = \zeros_R$.
\end{defi}  
\begin{defi}[Independent inlets]
The $p$ inlet streams are said to be independent if (i) the columns of  $\,\Wbf_{\inn}$  are linearly independent, i.e.\ $\rank{\Wbf_{\inn}}=p$, and (ii) there exists some finite time interval  $t \in [t_0,\,t_1]$
for which the inlet mass flowrate profiles $\ubf_{\inn}(t)$ are linearly independent, i.e.\ $\betabf^{\tr}\ubf_{\inn}(t) = 0$ $\Leftrightarrow$ $\betabf = \zeros_p$.
\end{defi}
 The  examples of independent reactions and  inlets can be found in Amrhein\citep{Amrhein98}.
\subsection{Transformation to extents}
Consider the homogeneous reaction system described by \eqrefn{sys1}. If $\rank{[ \Nbf^{\tr} \, \Wbf_{\inn} \, \nbf_0]} = R+p+1$, then $\mathcal{T}: \bMatsmall{\nbf \\ m} \rightarrow \bMatsmall{\xbf}$ such that, 

 \beali{
 \bMat{{\xbf}_\reac(t) \\ {\xbf}_\inn(t)  \\ {\lambda}(t) \\ {\xbf}_\inv(t)} =  \mathcal{T} \, \nbf(t).  \label{trans_forward1}
 }
Then, the dynamic model given in  \eqrefn{sys1} becomes
\begin{align}
\dot{\xbf}_{\reac}(t) &=\rbf_v(t) \,  - \omega(t) \,\xbf_{\reac}(t), &  \xbf_{\reac}(0) &=\zeros_R, \label{eq:main_trans_vessel_a}\\ 
\dot{\xbf}_{\inn}(t) &=\ubf_{\inn}(t)\,  -  \omega(t)\,\xbf_{\inn}(t),  &  \xbf_{\inn}(0) &=\zeros_p,  \label{eq:main_trans_vessel_b}\\ 
\dot{\lambda}(t) &=  - \omega(t)\,\lambda(t),  &  \lambda(0) &=1,\label{eq:main_trans_vessel_c}\\
\dot{\xbf}_{\inv}(t) &=  - \omega(t)\,\xbf_{\inv}(t),  & \xbf_{\inv}(0) &=\zeros_{S-\sigma}, 
\label{eq:main_trans_vessel_1}
\end{align}
where $\xbf_{\reac}(t)$ corresponds to the extent of reaction, $\xbf_{\inn}(t)$ corresponds to the extent of inlet, $\lambda(t)$ is a discounting variable for the initial conditions and $\xbf_{\inv}(t)$ is the extent of invariants that stays constant during the course of the reaction with $\sigma=S-R-p-1$.  The back transformation  $\mathcal{T}^{-1}: \bMatsmall{\xbf(t) } \rightarrow \bMatsmall{\nbf(t) \\ m(t)}$ can be given by:
\beali{
{\nbf}(t) & =  \Nbf^{\tr}{\xbf}_{\reac}(t)  +\Wbf_{\inn}{\xbf}_{\inn}(t) +\nbf_{0}\,{\lambda}(t), \\
m(t) & = \ones^{\tr}_p \xbf_{\inn}(t) + m_0 \lambda(t).
 \label{reconstruction_discount}
}
{\bf Remark} If $\rank{[ \Nbf^{\tr} \, \Wbf_{\inn} \, \nbf_0]} = S<R+p+1$, then the forward transformation does not hold, however, the back transformation in \eqrefn{reconstruction_discount} holds. Hence, the dynamic model in the extent domain can be used.  Further, the details of the linear transformation  in \eqrefn{trans_forward1} can be found in  Amrhein et al \cite{AmrheinBSB10} and Srinivasan et al \cite{SrinivasanBB13}. 
\subsection{Cooperative dynamical systems}

Consider the following nonnegative dynamical system,
\bea{
\dot\xbf=\fbf(\xbf,\ubf), \quad \xbf(0)=\xbf_0 \label{coopsys}
}
where $\xbf$ is  an $n$-dimensional  vector of states, $\fbf(\xbf,\ubf)$ the $n$-dimensional vector fields, and $\ubf$ the $p$-dimensional vector of inputs.  It will be assumed that  $\fbf(\xbf): \Rbb^n_{\geq 0}  \rightarrow \Rbb^n_{\geq 0}$ is locally Lipschitz with respect to $\xbf$.

\ \\
If the dynamic system described by \eqrefn{coopsys} is a cooperative system, then for the given sets of $k$ initial conditions such that $\xbf_1(0) \leq \xbf_2(0) \ldots \leq\xbf_k(0)$, the solution of this system will also follow  the inequalities $\xbf_1(t) \leq \xbf_2(t) \ldots \leq\xbf_k(t)$.  The operator $\geq$ or $\leq$ applied to matrices or vectors can be understood as a collection of inequalities applied to their elements.
The following lemma will give the conditions for the dynamical system in \eqrefn{coopsys} to be a cooperative system \citep{Smith95}:
\begin{lemma}\label{lemma:coop}
A given dynamical system in \eqrefn{coopsys} is a cooperative system, if   
$$\grad{f_j}{x_i} \geq 0 \quad \forall i\neq j, \quad \forall t\geq 0$$ 
\end{lemma}
Lemma~\ref{lemma:coop} states that the  non-diagonal elements of the Jacobian matrix of the system in \eqrefn{coopsys} have to be nonnegative for the system to be cooperative. In other words, the given system is said to be cooperative system, if the Jacobian matrix is a Metzler matrix \cite{Luenberger79}. In the next section, we will  derive  conditions under which  the system in the extents domain is cooperative system. \\

\section{Conditions for reaction systems to be cooperative system in extents domain}

Consider the system given in \eqrefn{eq:main_trans_vessel_1}. Let us represent the system as:
\begin{align}
\dot{\xbf}_{\reac}(t) &=\fbf_r, &\xbf_\reac(0)&=\zeros_R, \label{eq:main_trans_vessel_a}\\ 
\dot{\xbf}_{\inn}(t) &=\fbf_{\inn}, &  \xbf_{\inn}(0) &=\zeros_p,  \label{eq:main_trans_vessel_b}\\ 
\dot{\lambda}(t) &=  \fbf_{\lambda},  &  \lambda(0) &=1,\label{eq:main_trans_vessel_c}\\
\dot{\xbf}_{\inv} &=\fbf_{\inv},& \xbf_{\inv}(0) &=\zeros_{S-\sigma}. 
\label{eq:main_trans_vessel_1a}
\end{align}

We can then write the $(S \times S)$-dimensional the Jacobian  matrix $\Jbf_\xbf$ as follows:
\bea{
\Jbf_\xbf  =  \bMat{ \grad{\fbf_r}{\xbf_\reac} & \grad{\fbf_r}{\xbf_{\inn}} & \grad{\fbf_r}{\lambda} & \zeros_{R \times S-\sigma} \\  \zeros_{p \times R} & \grad{\fbf_{\inn}}{\xbf_{\inn}} & \grad{\fbf_{\inn}}{\lambda} & \zeros_{p \times S-\sigma} \\  \zeros_R^{\tr} & \grad{\fbf_{\lambda}}{\xbf_{\inn}} & \grad{\fbf_{\lambda}}{\lambda} &\zeros_{S-\sigma}^{\tr} \\  \zeros_{ S-\sigma \times R } & \zeros_{S-\sigma \times p} & \zeros_{S-\sigma} & \zeros_{S-\sigma \times S-\sigma}},
}
where
\begin{eqnarray*}
\bigg[\grad{\fbf_r}{\xbf_\reac}\bigg]_{R \times R}  &= & \grad{\rbf_v}{\xbf_\reac}-\frac{u_\out}{m}\Ibf_R \\
\bigg[\grad{\fbf_r}{\xbf_{\inn}}\bigg] _{R \times p} & = &  \grad{\rbf_v}{\xbf_\inn} +\frac{u_\out} {m^2}\xbf_\reac\ones_p^{\tr} \\
\bigg[\grad{\fbf_r}{\lambda}\bigg]_{R \times 1} & = & \grad{\rbf_v}{\lambda}+\frac{u_\out} {m^2} m_0\xbf_\reac \\ 
\bigg[\grad{\fbf_{\inn}}{\xbf_{\inn}}\bigg]_{p \times p}   & = & u_\out \bigg( \frac{1}{m^2}\xbf_\inn\ones_p^{\tr}-\Ibf_p/m \bigg) \\
\bigg[\grad{\fbf_{\inn}}{\lambda}\bigg]_{p \times 1}  & = & \frac{u_\out\, m_0,\xbf_\inn}{m^2}\ones_p \\
\bigg[\grad{\fbf_{\lambda}}{\xbf_{\inn}}\bigg]_{1 \times p}  & = & \frac{u_\out\, \lambda}{m^2}\ones_p^{\tr} \\
\bigg[\grad{\fbf_{\lambda}}{\lambda}\bigg]  & = & -u_\out\bigg (\frac{1}{m}-\frac{m_0\lambda}{m^2}\bigg).
\end{eqnarray*}
 According to Lemma~\ref{lemma:coop},  we need to investigate  the non-diagonal elements of $\Jbf_\xbf$ for a reaction system to be cooperative.

The extents of  flow are decoupled from the embedded reactions, and, hence, their cooperative nature can be studied independent of the extents of reaction.  The following proposition shows that the reaction invariants are cooperative in the extent domain.

\begin{proposition}\label{coopxinl}
The extents of flow (a subsystem defined by Eqs.~\eqref{eq:main_trans_vessel_b}-\eqref{eq:main_trans_vessel_c}) exhibit cooperative behavior.  
\end{proposition}\paragraph{Proof:}
The proof is divided into two parts. First, we show that $\xbf_{\inn}$ and $\lambda$ are strictly nonnegative functions . Then, we show that this non-negativity ensures that the extent of flow are always cooperative and satisfy the conditions in Lemma~ref{lemma:coop}.  Consider the dynamic equations of $\xbf_{\inn}$ and $\lambda$ (Eqs.~(12)--(13)):
\beali{
\dot{\zbf}(t) &=& \Qbf(t) - \frac{u_{\out}(t)}{m(t)} \zbf \qquad \zbf(0) \geq 0
}
where, $\zbf = \left[\begin{array}{c} \xbf_{\inn}  \\ \lambda \end{array}\right]$ and $\Qbf = \left[\begin{array}{c} \ubf_{\inn}  \\ 0 \end{array}\right]$. For constant $m$, the solution to this set of ODEs can be written as:
\beali{
\zbf(t) v(t) & = & v(0) \zbf(0) + \int\limits_0^t v(t) \Qbf(t)dt, \label{diff_eqsol}
}
where $v(t)$ is a positive function. Since $\zbf(0) \geq 0$ and $\Qbf(t) \geq 0$, the RHS is nonnegative in the interval [0,\,t], and hence, $\zbf(t) \geq \zeros$. 

Notice that since the mass of the reaction mixture $m>0$, the off-diagonal elements of the matrices $(\grad{\fbf_{\inn}}{\xbf_{\inn}})$, $(\grad{\fbf_{\inn}}{\lambda})$ and $(\grad{\fbf_{\lambda}}{\xbf_{\inn}})$ will always be non-negative for $t \geq 0$, making the extents of flow cooperative.  \qed

Proposition~\ref{coopxinl} establishes that the subsystem defined by the external material exchange (inlet and outlet flowrates) in the extent domain is always cooperative system. Next, the cooperative nature of the subsystem defined by the embedded reactions is studied. 
The elements of $\grad{\fbf_r}{\xbf_\reac}$, $\grad{\fbf_r}{\xbf_{\inn}}$, and $\grad{\fbf_r}{\lambda}$ of $\Jbf_\xbf$ need to be examined for  the cooperative behavior of the subsystem defined by the extents of reaction.   The following conditions must be satisfied for the extents of reaction to be a cooperative system:
\bea{
\grad{(V\,r_i)}{x_{\reac,j}} &\geq & 0,  \label{conditions_reaction1} \\ 
\grad{(V\,r_i)}{x_{\inn,l}}+\frac{u_\out} {m^2}x_{\reac,i} & \geq & 0 , \\  \label{conditions_reaction2}
\grad{(V\ r_i)}{\lambda}+\frac{u_\out} {m^2} m_0 x_{\reac,i}& \geq & 0  \label{conditions_reaction3}
}
$$\quad\forall i,j=1,\ldots,R,\,\, l=1,\ldots,p,\,\, i\neq j$$
From  Eqs.~\eqref{conditions_reaction1}--\eqref{conditions_reaction3}, it can be seen that the derivatives are  functions of the reaction rates. The rate of $i$th reaction is a function of concentrations, and it can be expressed by the following generic function:
\bea{
r_i & = & \frac {k_{f,i} \prod\limits^{S_{r,i}}_{e=1} c_e^{m_{i,e}}  - k_{b,i} \prod\limits^{S_{p,i}}_{k=1} c_k^{n_{i,k}}} {\prod\limits^{S_{r,i}}_{e=1} (a_{i,e} + c_e)^{d_{i,e}}}, \label{generalstuc} 
}
where $k_{f,i}$, $k_{b,i}$ and $a_{i,e}$ are nonnegative rate-constant parameters. \eqrefn{generalstuc} is assumed to be $C^{1}$ smooth function. Note that the mass-action  or Michaelis--Menten kinetics can be expressed with  \eqrefn{generalstuc}.  The denominator in \eqrefn{generalstuc} can alternatively expressed as $(a_{i,e} + c_e^{d_{i,e}})$ without loss of any generality.
The concentrations can be written as a function of the extents by using \eqrefn{reconstruction_discount} and volume of the reactor, 
\beali{
{\cbf}(t) & =  \frac{\Nbf^{\tr}{\xbf}_{\reac}(t)  +\Wbf_{\inn}{\xbf}_{\inn}(t) +\nbf_{0}\,{\lambda}(t)}{V(t)}. \label{conc_recon}}
For the constant volume system, Eqs.~\eqref{conditions_reaction1}--\eqref{conditions_reaction3} can be simplified to the following equations using   \eqrefn{conc_recon}:
\bea{
\grad{r_i}{\cbf}\Nbf^{\tr}_j &\geq & 0, \forall i,j=1,\ldots,R, i\neq j,  \label{eqreac}\\ 
\grad{r_i}{\cbf}\Wbf_{\inn,l} & \geq & 0, \forall l=1,\ldots,p, \label{eq_inlet}\\
\grad{r_i}{\cbf}\nbf_{0} & \geq & 0. \label{eq_ini} }
The above conditions can be understood as follows:  \eqrefn{eqreac} is the condition to be satisfied for the system to be cooperative with respect to the embedded reactions, while Eqs.~\eqref{eq_inlet}--\eqref{eq_ini} are related to   the operation modes  of the reaction system, for example, batch, semi-batch or continuous reactors (CSTR). \\
\ \\
 \textbf{Assumption A2:}
 $\grad{r_i}{c_e}$, $\forall e$=$1,\ldots,S_{r,i}$ and $\grad{r_i}{c_k}, \forall k=1,\ldots,S_{p,i}$ satisfy the following conditions for the reactants and products, respectively: \\
\beali{
\grad{r_i}{c_e} &\geq0,  \qquad \text{if}\,\, m_{i,e} \geq 0, n_{i,k} \geq 0  &\& & \quad  d_{i,e} = 0 \\
\grad{r_i}{c_k} &< 0,\qquad \text{if}\,\, m_{i,e} \geq 0, n_{i,k} > 0  &\& & \quad  d_{i,e} = 0  \\
\grad{r_i}{c_e} &> 0, \qquad \text{if}\,\,\qquad m_{i,e} >  d_{i,e}>0   &\&& \quad n_{i,k} \geq 0. 
 \label{drdc_cond}
}
We will next analyze in more detail the different scenarios for which the extents of reaction (a subsystem described by \eqrefn{eq:main_trans_vessel_a})  are cooperative in nature. 
\begin{proposition}\label{ReactionCom}
Consider a  system consisting $R$ reactions satisfying Assumptions A1 and A2.
For all the $ith$ and $jth$ reactions ($i\neq j$), if they have the same reactant set, i.e. $\mathcal{S}_i^R$ = $\mathcal{S}_j^R$ and produce different products in each reaction, i.e. $\mathcal{S}_i^P \neq \mathcal{S}_j^P$, such that $\mathcal{S}_i^P \cap \mathcal{S}_j^P = \phi$, these reactions are competitive in the extent domain.  
\end{proposition}\vspace{-0.5cm}\paragraph{Proof:} \eqrefn{eqreac}  can be written as follows:
\bea{
\grad{r_i}{x_{\reac,j}} =  \sum\limits_{k=1}^{S}\grad{r_i}{c_k} \nu_{k,j} \label{prop21}
}
Since  $\mathcal{S}_i^R$ = $\mathcal{S}_j^R$ and $\mathcal{S}_i^P \cap \mathcal{S}_j^P = \phi$, \eqrefn{prop21} reduces to:
\bea{
\grad{r_i}{x_{\reac,j}} =  \sum\limits_{e=1}^{S_{r,i}}\grad{r_i}{c_e} \nu_{e,j} \label{prop22}
}
From Assumption A2, $\grad{r_i}{c_e}>0$ and  $\nu_{e,j} < 0$ for the reactants, $\grad{r_i}{x_{\reac,j}}$ is strictly negative. Similarly, we can prove that $\grad{r_j}{x_{\reac,i}}$ is strictly negative. Hence, it is proven that the reactions are competitive in the extent domain.  \qed


\textbf{Illustrative example:} Consider the following reaction system.
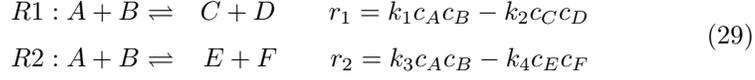
\beali{
R1: A + B &\rightleftharpoons& C + D \qquad r_1 = k_1 c_A c_B - k_2 c_C c_D \\
R2: A + B &\rightleftharpoons& E + F \qquad r_2 = k_3 c_A c_B - k_4 c_E c_F \\
}
Computing the derivative of R1 with respect to R2, we have 
\beali{
\grad{r_1}{x_{\reac,2}} &= -(k_1 c_B  + k_1 c_A) &\leq 0  \nonumber \\
\grad{r_2}{x_{\reac,1}} &= -(k_3 c_B  + k_3 c_A) &\leq 0 \nonumber\\
}
We can see that the system is competitive.   Proposition~\ref{ReactionCom} can also be applied to reach the same conclusion.  The above system can be made cooperative by a linear transformation. 
 By rewriting R2 as follows:
\beali{
E + F &\rightleftharpoons& A + B \qquad r_2 =  k_4 c_E c_F - k_3 c_A c_B \\
}
Then, the reaction system is cooperative as shown below:
\beali{
\grad{r_1}{x_{\reac,2}} &= k_1 c_B  + k_1 c_A  &\geq 0, \nonumber \\
\grad{r_2}{x_{\reac,1}} &= k_3 c_B  + k_3 c_A  &\geq 0. \nonumber
}

Next, the conditions under which the reaction set will be cooperative, are given. 
\begin{proposition}\label{coopprop}
If the reactant sets of  the reactions $i$ and $j$ are disjoint, (i.e). $\mathcal{S}_i^R \cap \mathcal{S}_j^R = \phi$, and Assumptions A1 and A2 hold, then the two reactions are cooperative.
\end{proposition}
\paragraph{Proof:}
The derivatives of the reaction rates with respect to the extents follows the same structure as \eqrefn{prop21} and \eqrefn{prop22}. Since the reactant sets are disjoint,  $\nu_{k,j}$ and $\nu_{e,i}$ are nonnegative. Hence,  the gradients are strictly non-negative, and  the reactions are cooperative. 

In practice, there are reaction systems the each set of reactant produces different products. This type of reactions are called series reactions. \qed 

\begin{corollary}
If Assumptions A1 and A2 hold, series reactions are always cooperative in the extent domain. 
\end{corollary}
\paragraph{Proof:}
Consider the reactions $i$ and $j$ in a system of series reactions. Without loss of generality, it is assumed that the reaction $j$ is followed by the reaction $i$. Then,  $\mathcal{S}_j^R = \mathcal{S}_i^P$, and hence, $\mathcal{S}_i^R \cap \mathcal{S}_j^R =\mathcal{S}_i^R \cap \mathcal{S}_i^P = \phi$.  Then from Proposition~\ref{coopprop}, the $i$th and $j$th reactions are cooperative in the extent domain. \qed  

\textbf{Illustrative example:} Consider the following series reaction system. 
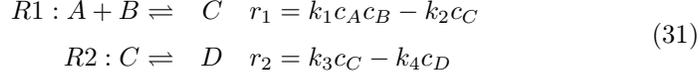
\beali{
R1: A + B &\rightleftharpoons& C \quad &r_1 = k_1 c_A c_B - k_2 c_C \\
R2: C &\rightleftharpoons & D \quad &r_2 = k_3 c_C - k_4 c_D \\
}
Writing the derivatives, 
\beali{
\grad{r_1}{x_{\reac,2}} &= -k_2 (-1) &\geq 0 \nonumber \\
\grad{r_2}{x_{\reac,1}} &= k_3   &\geq 0 \nonumber
}
The positive derivatives shows that the system is cooperative. \\
\ \\
Often, the reactants and products of one reaction react to produce new products via a new reaction. We will show that these kinds of reactions are cooperative provided they satisfy certain conditions. We define these kinds of system {\em conditionally cooperative} as described in the following proposition.

\begin{proposition}\label{prop4}
Consider the $ith$ and $jth$ reactions ($i\neq j$) of the reaction system. 
If a subset of the reactants of the $ith$ reaction react with a subset of the products of the $ith$ reaction in the reaction $j$, and Assumptions A1 and A2 hold, then the reactions are conditionally cooperative. 
\end{proposition}
\paragraph{Proof:}
In this kind of systems, $\mathcal{S}_j^R$ is a combination of reactants and products from the reaction $i$, (i.e.) $\mathcal{S}_j^R \subset (\mathcal{S}_i^R \cup \mathcal{S}_i^P)$ such that $\mathcal{S}_j^R \cap \mathcal{S}_i^R \neq \phi$ and $\mathcal{S}_j^R \cap \mathcal{S}_i^P \neq \phi$. Therefore, \eqrefn{eqreac} can be written as: 
\beali{
\grad{r_i}{x_{\reac,j}} = \sum\limits_{k=1}^{S_i^R}  \grad{r_i}{c_k} \nu_{k,j} - \sum\limits_{h=1}^{S_i^P}  \grad{r_i}{c_h} \nu_{h,j}
}
Here $\nu_{k,j}$ and $\nu_{h,j}$ are negative, leading rise to the following condition:
\beali{
	\grad{r_i}{x_{\reac,j}} =  \sum\limits_{h=1}^{S_i^P}  \grad{r_i}{c_h} |\nu_{h,j}| -  \sum\limits_{k=1}^{S_i^R}  \grad{r_i}{c_k} |\nu_{k,j}|\geq 0 \label{cc1}
}
Similarly, for the  reaction $j$ w.r.t the reaction $i$, we get the following condition: 
\beali{
\grad{r_j}{x_{\reac,i}} =  \sum\limits_{h=1}^{S_j^P}  \grad{r_i}{c_h} | \nu_{h,i} |-  \sum\limits_{k=1}^{S_j^R}  \grad{r_i}{c_k} |\nu_{k,i} |\geq 0 \label{cc2}
}
 Eqs.~\eqref{cc1}--\eqref{cc2} have to be satisfied for the dynamics of the $i$ and $j$ reactions  to be cooperative. Hence,  the system is conditionally cooperative. The violation of these conditions makes the system competitive.  \qed

\textbf{Illustrative example:} Consider the following reaction system. 
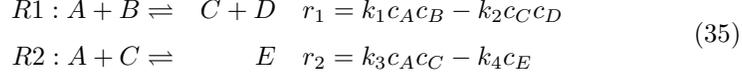
\beali{
R1: A + B &\rightleftharpoons& C + D \quad &r_1 = k_1 c_A c_B - k_2 c_C c_D \\
R2: A + C &\rightleftharpoons & E \quad &r_2 = k_3 c_A c_C - k_4 c_E \\
}
The off-diagonal elements are computed as follows.
\beali{
\grad{r_1}{x_{\reac,2}} &= k_1c_B (-1) + k_2 c_D  \nonumber \\
\grad{r_2}{x_{\reac,1}} &= k_3 c_C (-1) + k_3 c_A.  \nonumber
}
Here, $\mathcal{S}_i^R=\{A,B\}$, $\mathcal{S}_i^P=\{C,D\}$, and $\mathcal{S}_j^R=\{A,C\}$. Since $\mathcal{S}_j^R \subset \mathcal{S}_i^R \cup \mathcal{S}_i^P$, 
 the system is  a conditionally cooperative system in the extent domain according to  Proposition~\ref{prop4}.  The conditions are as follows: $k_2c_D \geq k_1c_B$ and $k_3c_A \geq k_3c_C$. \\
\ \\
So far, we have investigated conditions under which the embedded reactions are cooperative or competitive in the extent domain. Next, we will investigate how the operation modes  such as the inlet compositions and the initial conditions affect with the embedded reactions in the extent domain. The next propositions discuss the influence of the inlet compositions and the initial conditions on the cooperative behavior of the embedded reactions with the extents of flow. 

\begin{proposition}[Inlet Compositions]
If the $jth$ inlet stream feeds only the reactants of the $ith$ reaction, then,  term $\grad{\fbf_{\reac,i}}{\xbf_{\inn,j}}$ is always positive.
\end{proposition}
\paragraph{Proof:}
Assume that the $jth$ stream feeds only the reactants. Since the elements of $\Wbf_{\inn,j}$ contains weight fractions ($w_{\inn,j}$), they are either zero or positive. Now, we can write
\bea{
\grad{\fbf_{\reac,i}}{\xbf_{\inn,j}} = \sum\limits_{j=1}^{S_{r,i}}\grad{r_i}{c_j}w_{\inn,j} \label{reac_inlet1}}
Since, the inlet contains only the reactants, $\grad{r_i}{c_j}$ is positive by \eqrefn{drdc_cond}. This makes \eqrefn{reac_inlet1} always greater than zero. \qed

\textbf{Note:} We can also show that:
\begin{itemize}
\item If the $jth$ inlet stream feeds only products of the $ith$ reaction, the term $\grad{\fbf_{\reac,i}}{\xbf_{\inn,j}}$ is always negative.
\item If the $jth$ inlet stream feeds both reactants and products of the $ith$ reaction, the term $\grad{\fbf_{\reac,i}}{\xbf_{\inn,j}}$ could be either nonnegative or negative.
\item In batch mode of operation, $\grad{\fbf_{\reac,i}}{\xbf_{\inn,j}}=0$.
\end{itemize}
\begin{proposition}[Initial Conditions] \label{ICcoop}
If  the reaction mixture contains the reactants of the $ith$ reaction  initially, then the term $\grad{\fbf_{\reac,i}}{\lambda}$ is always positive. 
\end{proposition}
\paragraph{Proof:}
We know that the initial conditions $\nbf_0$ are either positive or zero. We can write $\grad{\fbf_{r,i}}{\lambda}$ as: 
\bea{
\grad{\fbf_{\reac,i}}{\lambda} = \sum\limits_{j=1}^{S_{r,i}}\grad{r_i}{c_j}n_{0,j} \label{reac_initial1}}
Since $\grad{r_i}{c_j}$ is always positive for the reactants, we have a strictly positive function. \qed

\textbf{Note:} Similar to the conditions for the inlet compositions, we can also show that: 
\begin{itemize}
\item If  the reaction mixture contains the products of  the reaction $i$ initially, the term $\grad{\fbf_{\reac,i}}{\lambda}$ is always negative.
\item The presence of both reactants and products of a reaction $i$, makes the term  $\grad{\fbf_{\reac,i}}{\lambda}$ either nonnegative or negative.
\item In batch or semi-batch modes of operation, $\grad{\fbf_{\reac,i}}{\lambda}=0$.
\end{itemize}
\section{Examples}
Two examples are considered in this section to demonstrate various results. The first example is from field of control of bio-reactors \cite{RapaportD05}, and the second example is from systems biology\cite{AngeliLS10}. 

{\bf Example 1:} Consider a reaction in which the substrate $S$ forms a biomass $X$ in a stirred tank fed batch reactor  as follows\cite{RapaportD05}: $\frac{1}{k}S \rightarrow X$.  The mass balance equations in terms of concentrations can be written as follows: 
\beali{
\dot{S}(t) & =  -\frac{1}{k} \mu(S) X + D S_{\inn} - DS, \quad S(0)=S_0, \\
\dot{X}(t) & =  \mu(S) X  - DX, \quad X(0)=0. \label{Simple-bio}
}
where $k$ is the yield coefficient, $\mu(S)=\mu_{\max} \frac{S}{K_s + S}$ the specific growth rate, $D$ is the dilution rate, $S_{\inn}$ is the inlet substrate concentration. 
 Rapaport and Dochain \cite{RapaportD05} have showed that the system is competitive in the concentration domain. 
We can show that it is cooperative in the extent domain as follows.  The model equations in terms of extents can be written as follows:
\beali{
\dot{\xbf}_{\reac} &= \mu(S) X - D \,  \xbf_{\reac} \\
\dot{\xbf}_{\inn} &= D - D \xbf_{\inn} \\
\dot{\lambda} & = -D \lambda 
}
The concentrations can be written as a function of the extents: 
\beali{
S &= \frac{-1}{k} \xbf_{\reac} + S_{\inn} \xbf_{\inn} + S_0 \lambda \\
X &= \xbf_{\reac} + X_0 \lambda
}
According to Proposition~\ref{coopxinl}, the extents of flow are cooperative. Hence,
in order to check the  system being cooperative, we need to make sure that the off-diagonal elements $\grad{\xbf_{\reac}}{\xbf_{\inn}}$ and $\grad{\xbf_{\reac}}{\lambda}$ are non-negative. Since we feed only the reactant, the term $\grad{\xbf_{\reac}}{\xbf_{\inn}}$ is positive (Proposition 5). Since the initial concentration of biomass $X_0$ is zero, the term $\grad{\xbf_{\reac}}{\lambda}$ will also be positive  (Proposition 6). Hence, the system is cooperative in the extent domain. \\
\ \\
{\bf Example 2:} Consider a reaction system with the substrate $S_1$ which is converted into a product $S_2$ in the  presence of an enzyme $E$\cite{AngeliLS10}. Then the species $S_2$ produces the substrate $S_1$ due to the presence of another enzyme $F$.   The embedded reactions are as follows: 
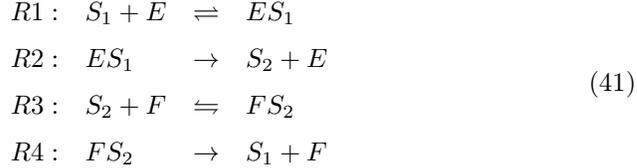
\beali{
R1: &\quad S_1 + E &\rightleftharpoons \quad& ES_1 & \\
R2: &\quad ES_1  &\rightarrow \quad& S_2 + E  &\\
R3: &\quad S_2 + F  &\leftrightharpoons \quad& FS_2& \\
R4: &\quad FS_2 &\rightarrow  \quad& S_1 + F  &
}
\ \\
Here, $\mathcal{S}_1^R= \{ S_1,\, E\}$, $\mathcal{S}_2^R= \{ ES_1\}$, $\mathcal{S}_3^R= \{ S_2,\, F\}$, and $\mathcal{S}_4^R= \{ FS_2\}$. Since $\mathcal{S}_i^R \cap \mathcal{S}_j^R = \phi$, $\forall i,j=1,2,3,4,\, i\neq j$, according to Proposition~\ref{coopprop}, the  embedded reactions are cooperative.  Further, since the operation mode is batch in the nature, $\grad{\fbf_{\reac,i}}{\lambda}=0$. Hence, the reaction system is cooperative system.
\section{Conclusions}
The paper has studied the cooperative behavior of open homogeneous reaction systems in the extent domain. It has been shown that  for open reaction systems, the cooperative conditions can be classified into two parts. The first part deals with the embedded reactions and the second part deals with the operation mode. It has been shown that   the extents of flow (the dynamics of operation mode) exhibit cooperative behavior in the extent domain. Further, it has clearly come out that the embedded reactions are necessary condition for  be cooperative behavior of open reaction systems. Various conditions under which the embedded reactions exhibit cooperative or competitive behavior have been derived. Further, the effect of operation mode on the cooperative behavior of the embedded reactions have been studied. 
This paper proposes  a set of  simple  to check  and physical meaningful conditions for cooperative or competitive behavior of open reaction systems in the extent domain.  The results have been demonstrated via two examples from  the fields of   process control (Example 1), and systems biology (Example 2). 
\section*{Acknowledgment}
 The financial support to NB from Department of Science $\&$ Technology, India through INSPIRE Faculty Fellowship is acknowledged.

\end{document}